# Liquid-fueled oblique detonation waves induced by reactive and non-reactive transverse liquid jets

Wenhao Wang[a,b,c], Zongmin Hu[a,c,*], Peng Zhang[b,*]

[a]*State Key Laboratory of High-temperature Gas Dynamics (LHD), Institute of Mechanics, Chinese Academy of Sciences, Beijing, 100190, China*
[b]*Department of Mechanical Engineering, City University of Hong Kong, Kowloon Tong, Kowloon, 999077, Hong Kong*
[c]*School of Engineering Science, University of Chinese Academy of Sciences, Beijing, 100049, China*

___

**Abstract**

*This computational study demonstrates the formation of liquid-fueled oblique detonation waves (ODWs) induced by a liquid transverse jet, which is either reactive or non-reactive. The study employs an in-house two-phase supersonic reactive flow solver based on the rhocentralfoam framework of OpenFOAM. The findings emphasize the essential role of transverse jets in enabling successful ODW formation under conditions where detonation would otherwise fail. Specifically, the jet-inflow momentum ratio significantly influences the mechanisms of ODW formation. At lower momentum ratios, the oblique shock wave (OSW) induced by the jet is insufficient to directly initiate detonation. Instead, the atomized n-heptane jet increases the local fuel mass fraction, promoting low- and intermediate-temperature chemical reactions, which eventually lead to detonation. At higher momentum ratios, the OSW generated by the transverse jet is sufficiently strong to directly trigger detonation through intermediate-temperature chemistry, with the jet acting primarily as a combustion stabilizer rather than directly enhancing combustion. Comparative studies with non-reactive jets and wedge-strip configurations demonstrate that at higher momentum ratios, the dominant mechanism is the physical blocking effect of the jet, which generates a strong OSW capable of initiating detonation.*

*Keywords:* Jet-induced oblique detonation waves; Liquid-fueled detonation; Transverse liquid jet; OpenFOAM.

___

*Corresponding author.

Email addresses: huzm@imech.ac.cn, (Z. Hu); penzhang@cityu.edu.hk, (P. Zhang).

**Novelty and significance statement**

The novelty of this work lies in presenting the first computational investigation of liquid-fueled oblique detonation waves (ODWs) induced by transverse liquid jets in an oblique detonation wave engine, where the jet can be either reactive or non-reactive. Using an in-house two-phase supersonic chemically reacting flow solver based on OpenFOAM, the study explores the impact of jet-inflow momentum ratios and jet types on ODW formation. This work is significant because it identifies two distinct mechanisms for ODW formation: at lower momentum ratios, the shock wave induced by the jet is insufficient to initiate ODWs directly. Instead, it promotes low- and intermediate-temperature chemical reactions, facilitating the transition to detonation. At higher momentum ratios, the reactive jet behaves similarly to the non-reactive jet and on-wedge strip, generating a strong oblique shock wave that acts as a physical barrier, directly triggering detonation. These findings offer critical insights into controlling liquid-fueled ODWs in combustors.

**Nomenclature**

The nomenclature table is provided in the Supplementary Material.

## 1. Introduction

An oblique detonation wave (ODW) is an oblique shock wave (OSW) followed by rapid combustion [1-3]. The oblique detonation wave engine (ODWE) offers a high thermal cycle efficiency and a simple configuration, making it a promising candidate for hypersonic propulsion [4-7]. Among the various fuel options for ODWE, liquid hydrocarbon fuels are preferred due to their high energy density, high hydrogen content per unit volume, and ease of storage [8]. However, these fuels present challenges for forming and stabilizing a standing ODW in the engine due to their complex atomization process and long ignition delay time [9]. Therefore, understanding the formation of liquid-fueled ODWs and developing methods to control their formation is crucial for engine design and optimization.

The ignition and formation of ODWs is a critical issue for the successful operation of ODWE [10]. Considerable research has been dedicated to enhancing the formation of gaseous ODWs, particularly hydrogen-fueled ODWs, with various methods proposed [9, 11, 12]. Xin et al. [12] introduced a laser-heating hot spot, demonstrating its effectiveness in promoting the accelerated initiation of the ODW. They also explored the influence of hot-spot temperature, size, and position on the ODW initiation point. Sun et al. [13, 14] proposed a proportional controller for adjusting wedge angles, ensuring that the autoignition point remains at a desired location while stabilizing the ODW. Additionally, they performed computational investigations into the evolution of the ODW structure under unsteady inflow conditions and analyzed the function of the proportional controller. Qin et al. [15] introduced a wedge with a stepped surface configuration to trigger the ODW, discovering that this method effectively forms the ODW through a compression–expansion–compression cycle.

Among these techniques, the transverse jet method has garnered the most attention due to its flexibility in controlling ODW formation. Han et al. [11] conducted computational studies on cold-jet-induced hydrogen-fueled ODWs using in-house codes, analyzing how variations in jet-inflow momentum ratios and wedge angles affect the wave structure and thrust potential. Wang et al. [16] explored the formation of hydrogen-fueled ODWs induced by a hot jet, demonstrating that the jet can function as a gaseous wedge, inducing an OSW and thereby reducing the initiation length of ODWs. Li et al. [17] and Qin et al. [18] also utilized a hot jet to actively control the initiation length of the ODW. Yao et al. [19] computationally investigated a jet-wedge combinatorial method based on OpenFOAM, positioning the jet upstream of the wedge to enhance the ignition process. Fan et al. [20] applied the sweeping jet technique to the wedge, finding that it reduced the initiation length by 25% compared to scenarios without the jet.

Forming liquid-fueled ODWs presents significant challenges compared to gas-fueled ODWs, primarily due to the complex gas-liquid two-phase interaction. Despite the growing interest in gaseous detonation waves, the formation and enhancement methods for liquid-fueled ODWs have not been as extensively explored. This disparity stems from the intricate coupling between the liquid fuel and gas phase, which complicates the detonation process. Ren et al. [21] conducted a numerical investigation into wedge-induced ODWs in two-phase kerosene-air mixtures, employing a hybrid Eulerian-Lagrangian method combined with a particle-in-cell approach to account for the two-way coupling between phases. Their work highlighted significant differences in the initiation and stabilization characteristics between liquid-fueled and gaseous detonation waves. In follow-up studies, Ren et al. [22, 23] examined the effects of equivalence ratio and inflow pressure oscillations on ODW formation and stabilization. Zhang et al. [24] developed a two-phase supersonic solver named *RYrhoCentralFoam* based on the OpenFOAM platform for two-phase detonation. Building upon this tool, Guo et al. [25] numerically investigated wedge-induced ODWs in water mist environments, revealing the influence of mist on the flow field and chemical structure within the induction zone. They further explored the autoignition and transition behavior of n-heptane droplet/vapor/air mixtures behind an OSW using a $1 \times 1 \times 1$ mm³ closed reactor [26]. Teng et al. [27, 28] conducted simulations of ODWs in partially pre-vaporized n-heptane sprays. They observed that the initiation lengths of oblique detonation waves initially increased and then decreased with the droplet diameter, and they also identified the unsteady behavior of spray detonation.

Han et al. [9] conducted the pioneering liquid-fueled ODWE experiment in a hypersonic shock tunnel, utilizing an on-wedge trip to control ODW formation within the combustor. Their results confirmed that the on-wedge trip effectively facilitated ODW formation under liquid-fueled conditions. Building on this work, the authors computationally reproduced the experiments qualitatively, investigating the impacts of droplet breakup models, gas-liquid ratios, and on-wedge trips on forming liquid-fueled ODWs in a realistic combustor environment [29]. These findings emphasized the crucial role of ignition control methods in the successful formation of ODWs.

Based on the above discussions, it can be inferred that the formation of liquid-fueled ODWs presents significant challenges. Furthermore, most existing research on liquid-fueled ODWs primarily focuses on the initiation and unstable behavior of the detonation waves, with limited attention given to methods for enhancing the formation of liquid-fueled ODWs. Therefore, this study computationally demonstrates, for the first time, the formation of liquid n-heptane-fueled ODWs over wedges induced by various types of liquid jets, both reactive and non-reactive. It investigates the ODW formation behavior and examines the impact of jet types and momentum ratios on these processes, while also interpreting the formation mechanism through chemical kinetics analysis.

The computational methodology is described in Section 2. Section 2.1 presents the governing equations and associated sub-models, followed by the computational specifications in Section 2.2. A mesh-independence study is conducted in Section 2.3. Section 3 details the results and analysis, with Section 3.1 focusing on the phenomenology of liquid n-heptane transverse-jet-induced formation of ODWs. Section 3.2 explores the effects of jet-inflow momentum ratios while Section 3.3 examines the chemical kinetics analysis at varying momentum ratios. Section 3.4 compares the differences in ODW formation between reactive and non-reactive transverse liquid jets, as well as the on-wedge strip. Section 3.5 analyzes the mechanisms of ODW formation with reactive and non-reactive transverse liquid jets, and an on-wedge strip at high momentum ratios. Finally, Section 4 presents the conclusions and future research directions.

## 2. Computational methodology

### 2.1 Governing equations and sub-models

The present study employs an Eulerian-Lagrangian approach to simulate the two-phase compressible chemically reactive flow. The gas phase is modeled using an Eulerian framework, solving the compressible Navier-Stokes (N-S) equations for multi-component reactive flows, as given by:

$$\frac{\partial \rho}{\partial t} + \nabla \cdot (\rho \mathbf{u}) = S_m, \quad (1)$$

$$\frac{\partial (\rho \mathbf{u})}{\partial t} + \nabla \cdot [\mathbf{u}(\rho \mathbf{u})] + \nabla p - \nabla \cdot \boldsymbol{\tau} = \mathbf{S}_F, \quad (2)$$

$$\frac{\partial (\rho E)}{\partial t} + \nabla \cdot [\mathbf{u}(\rho E)] + \nabla \cdot (\mathbf{u}p) + \nabla \cdot \mathbf{q} - \nabla \cdot (\boldsymbol{\tau} \cdot \mathbf{u})$$
$$= S_e, \quad (3)$$

$$\frac{\partial (\rho Y_i)}{\partial t} + \nabla \cdot [\mathbf{u}(\rho Y_i)] + \nabla \cdot [-D\nabla(\rho Y_i)]$$
$$= \dot{\omega}_i + S_{s,i}, (i = 1,2,\dots,ns-1). \quad (4)$$

In Equ. (1) - (4), the variables $\rho$, $\mathbf{u}$, and $p$ represent the gas density, velocity, and pressure, respectively. The pressure $p$ satisfies the ideal gas law $p = \rho RT$, where $T$ is the gas temperature, and $R$ is the specific gas constant, calculated as $R = R_u \sum_{i=1}^{ns}(Y_i/MW_i)$. Here, $R_u = 8.314 \, J/(mol \cdot K)$ is the universal gas constant, and $MW_i$ is the molecular weight of the $i$-th species. $ns$ is the number of species, and $Y_1, \dots, Y_{ns-1}$ are the mass fractions of each species. The deviatoric stress tensor $\boldsymbol{\tau}$ is expressed as $\boldsymbol{\tau} = \mu[\nabla \mathbf{u} + (\nabla \mathbf{u})^T - 2(\nabla \cdot \mathbf{u})\mathbf{I}/3]$, where $\mu$ is the dynamic viscosity, calculated using Sutherland's law. $E = e + |\mathbf{u}|^2/2$ is the total energy, where $e$ is the internal energy. $\mathbf{q}$ is the diffusive heat flux calculated by Fourier's law as $\mathbf{q} = -k\nabla T$, where $k$ is the thermal conductivity. $D$ is the density-weighted diffusion coefficient, which can be computed using the unity Lewis number assumption as $D = k/\rho C_p$, where $C_p$ is the heat capacity at constant pressure given by $C_p = \sum_{i=1}^{ns} Y_i C_{p,i}$. The net production rate of the $i$-th species is denoted by $\dot{\omega}_i$. Specifically, a skeletal reaction mechanism comprising 44 species and 112 reactions is employed to model the chemical kinetics of n-heptane detonation [30]. This mechanism has been extensively validated against experimental data and is proven effective in reproducing key detonation characteristics in n-heptane-fueled systems [31-35]. The source terms $S_m$, $\mathbf{S}_F$, $S_e$, and $S_{s,i}$ originate from the liquid phase and will be further explained in the following text.

The liquid phase is modeled using a Lagrangian framework, with the Lagrangian Particle Tracking (LPT) method employed to track the motion liquid droplets, as well as their mass and temperature variations. The governing equations are given by:

$$\frac{dm_p}{dt} = \dot{m}_p, \quad (5)$$

$$\frac{d\mathbf{u}_p}{dt} = \frac{\mathbf{F}_p}{m_p}, \quad (6)$$

$$c_p \frac{dT_p}{dt} = \frac{\dot{Q}_c + \dot{Q}_l}{m_p}. \quad (7)$$

where $m_p$, $\mathbf{u}_p$, and $T_p$ represent the mass, velocity, and temperature of each droplet, respectively. The droplet mass is calculated as $m_p = \pi \rho_p D_p^3/6$ under the assumption that the droplet is spherical, where $\rho_p$ and $D_p$ are the density and diameter of the droplet. $c_p$ is the constant-pressure heat capacity of the liquid. The right-hand side terms of Equ. (5) – (7) are calculated using several sub-models.

For droplet mass transfer in Equ. (5), the evaporation model presented by Abramzon and Sirignano [36] is applied under the quasi-steady evaporation assumption. The droplet evaporation rate is expressed as $\dot{m}_p = -\pi D_p Sh \overline{D}_f \rho_s \ln(1 + B_M)$, where $\rho_s$ denotes the fuel vapor density at the surface of the droplet, and it is determined by the Clasius-Clapeyron formula $\rho_s = p_s MW_m/RT_s$, where $p_s$ is calculated using a polynomial function of $T_s$ [37], and the droplet surface is estimated by $T_s = (T + 2T_p)/3$. $\overline{D}_f$ is the average binary diffusion coefficient of the gas mixture in the film. The Sherwood number ($Sh$) is modelled as $Sh = 2.0 + 0.6 Re_p^{\frac{1}{2}} Sc^{\frac{1}{3}}$, where $Re_p = \rho_p D_p |\mathbf{u}_p - \mathbf{u}|/\mu$ is the Reynolds number of the droplet, and $Sc = \nu/\overline{D}_f$ is the Schmidt number. $\nu = \mu/\rho$ is the kinematic viscosity coefficient. The mass transfer number $B_M$ is given by $B_M = (Y_s - Y_g)/(1 - Y_s)$, where $Y_s$ and $Y_g$ are the mass fractions of fuel vapor at the droplets' surface and surrounding gas, respectively. The value of $Y_s$ is computed as $Y_s = MW_p X_s/\sum_i (X_i MW_i)$, where $MW_p$ is the molecular weight of the fuel vapor, and $X_s$ represents the mole fraction of the fuel vapor on the droplet surface. $X_s$ can be calculated by the Raoult's law as $X_s = X_m p_{sat}/p$, where $X_m$ is the mole fraction of the condensed species in the liquid phase, and $p_{sat}$ is the saturated pressure of the liquid fuel. The value of $p_{sat}$ is determined using a polynomial function of $T_p$ [37].

For the droplet force in Equ. (6), it is calculated using the sphere drag model as $\mathbf{F}_p = -(18\mu/\rho_p D_p^2)(C_d Re_p/24)m_p(\mathbf{u}_p - \mathbf{u})$, where $C_d$ is the drag coefficient [38]. For heat transfer between the gas and liquid phase in Equ. (7), the convective heat transfer rate is expressed as $\dot{Q}_c = h_c A_p (T - T_p)$, in which $A_p = \pi D_p^2$ is the surface area of the droplet, and $h_c = Nu \cdot k / D_p$ is the convective heat transfer coefficient. Here, $Nu$ is the Nusselt number, which is calculated using the Ranz-Marshall model [39] as $Nu = 2.0 + 0.6 Re_p^{\frac{1}{2}} Pr^{\frac{1}{3}}$, where $Pr = c_p \mu / k$ is the Prandtl number. The droplet evaporation heat transfer rate is $\dot{Q}_l = -\dot{m}_p h_l(T_p)$, where $h_l(T_p)$ is the latent heat of droplet at temperature $T_p$. For droplet breakup, the Pilch-Erdman model is utilized [40], which has been extensively applied in liquid-fueled detonation simulations [27, 34, 35, 41, 42].

To capture the interaction between the gas and liquid phases, a two-way coupling approach is employed using the Particle-Source-In-Cell (PSI-CELL) method [43]. The source terms in each calculation cell with cell volume $V_c$ and droplet particle number $N_p$ are expressed as follows,

$$S_m = -\frac{1}{V_c} \sum_1^{N_p} \dot{m}_p, \tag{8}$$

$$\boldsymbol{S}_F = -\frac{1}{V_c} \sum_1^{N_p} \boldsymbol{F}_p, \tag{9}$$

$$S_e = -\frac{1}{V_c} \sum_1^{N_p} (\dot{Q}_c + \dot{Q}_l), \tag{10}$$

$$S_{s,i} = \begin{cases} -\frac{1}{V_c} \sum_1^{N_p} \dot{m}_p, & \text{for condensed species} \\ 0, & \text{otherwise} \end{cases}. \tag{11}$$

In the present study, the two condensed species considered are $C_7H_{16}$ (n-heptane) and $H_2O$ (water), which are also the species present in both the reactive and non-reactive jets.

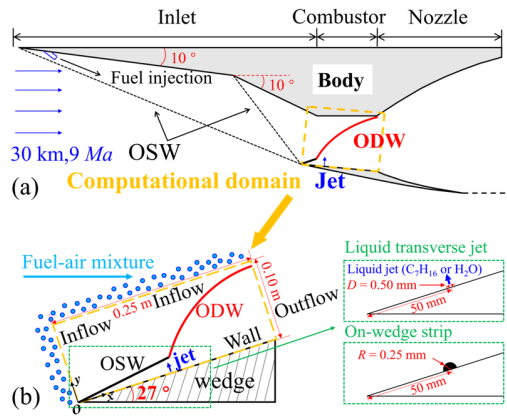

Fig. 1. Schematics of (a) the configuration of an external-injection ODWE and (b) the detailed geometry of the computational domain.

## 2.2 Computational specifications

This study employs a simplified combustor model of a pre-injection ODWE to investigate the formation of oblique detonation waves triggered by a liquid transverse jet. The ODWE is engineered to operate at Mach 9 in conditions consistent with an altitude of 30 km, with a two-stage compression inlet illustrated in Fig. 1(a). The inflow is compressed via two 10° wedges before entering the combustor. Following the approach by Teng et al. [27], the computational domain is depicted within the orange dashed box, where the combustor is simplified to a single wedge with a 27° angle, as shown in Fig. 1(b). The length of the computational domain is 0.25 m, and the height is 0.1 m, which are similar to the dimensions of a typical ODWE combustor [10, 44].

Regarding the boundary conditions, the inflow parameters for the computational domain are determined using shock relations, yielding a pressure of $p_1 = 2.855 \times 10^4$ Pa, a temperature of $T_1 = 697.0$ K, and a velocity of $u_1 = 2535.0 \; m/s$. The n-heptane droplets are assumed to be randomly distributed in the inflow air, with a droplet diameter of 10 μm, a temperature of 300 K, and a stoichiometric ratio of unity. The number of droplet parcels ranges from $5 \times 10^4$ to $1 \times 10^5$, with each droplet parcel containing 10 droplet particles, a quantity that strikes a balance between computational accuracy and efficiency, and is consistent with previous studies addressing two-phase detonation problems [35, 45]. As for the effect of pre-evaporation, the droplets travel a distance of approximately 0.1 m before encountering the OSW, with flight velocities around $2 \times 10^3$ m/s. Consequently, the flight time is approximately 0.05 ms. According to the $d^2$ law and experimental evaporation data for n-heptane droplets [46], the evaporation time of the n-heptane droplets under these conditions is approximately 0.5 ms. Therefore, the effect of pre-evaporation is negligible and does not significantly affect the results. Moreover, despite the temperature variations in the incoming gas caused by droplet evaporation, the ignition mechanism in this study is primarily governed by chemical and physical processes near the wall. The streamlines in this region show that both the incoming flow and droplets pass directly through the OSW. As a result, the impact of gas temperature changes due to droplet evaporation on the ignition process can be ignored.

For the outflow boundary shown in Fig. 1(b), zero-gradient Neumann boundary conditions are applied for pressure, temperature, and velocity, as the outflow is supersonic. For the wall boundary, a slip-reflecting boundary condition is employed. Although the boundary layer may interact with the jet and influence the flow structure, this preliminary study primarily focuses on the competition between physical and chemical ODW formation mechanisms. Given the thin boundary layer resulting from the high Reynolds number, its effect is neglected in this study and will be addressed in future work. This approach is consistent with previous studies [27, 35, 41, 47].

Two formation control strategies are employed to ensure successful ignition, as indicated by the green dotted box in Fig. 1(b). The first method involves the liquid transverse jet injector, positioned 50 mm downstream of the wedge's leading edge with a diameter of 0.5 mm. Two types of jets are used: a reactive jet ($C_7H_{16}$) and a non-reactive jet ($H_2O$). The jet velocity is regulated by the jet-inflow momentum ratio $J$, defined as

$$J = \frac{\rho_{jet} u_{jet}^2}{\rho_2 u_2^2}, \tag{12}$$

where $\rho_{jet}$ and $u_{jet}$ denote the density and velocity of the liquid jet, respectively, while $\rho_2$ and $u_2$ represent the density and velocity of the post-shock flow after the wedge-induced OSW. In this study, the primary $J$ values considered for the n-heptane jet are 0.1, 0.5, 1.0, and 2.0, which correspond to jet velocities of 19.24 m/s, 43.01 m/s, 60.83 m/s, and 86.02 m/s, respectively. For the water jet, the primary $J$ values considered are 0.1, 0.5, and 1.0, with corresponding velocities of 15.92 m/s, 35.60 m/s, and 50.35 m/s. For the jet, the Rosin-Rammler (R-R) distribution model is employed to substitute the primary atomization process and establish the droplet diameter distribution, with an average droplet diameter set to 10 μm. The second method is the on-wedge strip, which is also positioned 50 mm from the leading edge of the wedge and has a diameter of 0.5 mm.

The present study employs a recently developed open-source two-phase supersonic reactive flow solver, which was proposed by the authors in a previous study [48]. This solver integrates chemical reactions, multi-component transport, and a liquid-phase Lagrangian solver into the *rhocentralfoam* framework of OpenFOAM V7 [49]. It has been rigorously validated against theoretical predictions and experimental data for gaseous and liquid-fueled ODW problems, demonstrating accuracy and reliability [29, 48]. The first-order Euler scheme is used for the temporal discretization, and the "Gauss-Limited linear" scheme is used for spatial derivatives. The CFL number is set as 0.3 for the gas flow. To resolve shock waves and discontinuities in the detonation structure effectively, the KNP scheme is utilized [50].

### 2.3 Mesh-independence study

To verify the mesh independence of the present two-phase ODW simulations, a series of simulations were conducted using varying mesh sizes for the jet-induced liquid-fueled ODW problem involving an n-heptane liquid jet with a momentum ratio of unity. Four mesh resolutions were considered: 0.15 mm, 0.2 mm, 0.4 mm, and 0.8 mm, corresponding to total cell counts of $1.11\times10^6$, $6.25\times10^5$, $1.56\times10^5$, and $3.91\times10^4$, respectively. Temperature and pressure profiles along a streamline passing through the point ($x = 0$, $y = 0.02$ m) at $t = 0.1$ ms were extracted and plotted for comparison, as shown in Fig. 2.

The results indicate that the temperature and pressure curves for the two finest meshes (0.15 mm and 0.2 mm) are highly consistent, particularly in regions near the pressure and temperature peaks, demonstrating adequate resolution of the detonation wave structure. For the two coarser meshes (0.4 mm and 0.8 mm), notable deviations are observed, with the discrepancies becoming more pronounced for the coarsest mesh (0.8 mm). These findings demonstrate that the finest mesh (0.15 mm) and the second finest mesh (0.2 mm) provide sufficient accuracy for resolving the detonation wave structure. Considering the balance between accuracy and computational cost, the 0.2 mm mesh is selected for all subsequent simulations.

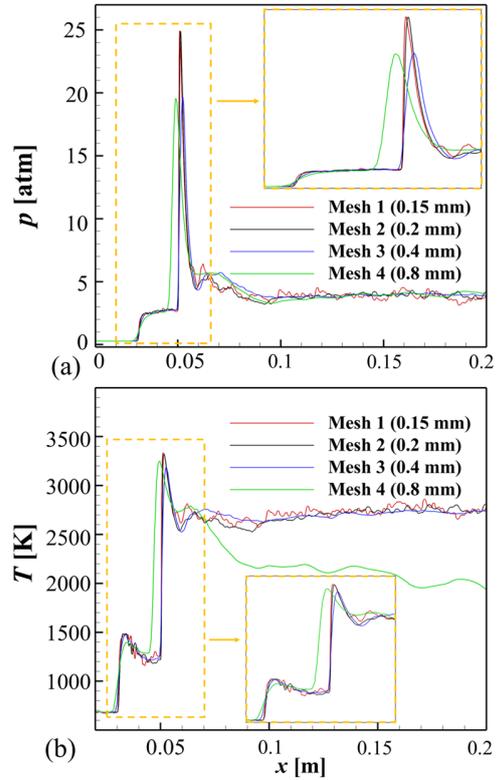

Fig. 2 The flow parameters along streamlines that pass through the same point ($x = 0$, $y = 0.02$ m) for cases with n-heptane vapor/air inflow across various mesh sizes: (a) Pressure curves, (b) Temperature curves.

## 3. Results and Discussion

### 3.1 Phenomenology of liquid n-heptane transverse-jet-induced formation of ODWs

To investigate the effect of a liquid n-heptane transverse jet on the formation of liquid-fueled ODWs, three representative cases with varying inflow and jet configurations are analyzed: (a) a pure n-heptane vapor/air mixture inflow without a transverse jet, (b) an n-heptane droplet/air mixture inflow without a transverse jet, and (c) an n-heptane droplet/air mixture inflow with a transverse jet. Figure 3 illustrates the flow fields of these three cases, with numerical Schlieren images in the first row and temperature contours in the second row. To better visualize the flow parameters in the ODW formation process, detailed profiles of key thermodynamic and chemical variables along streamlines near the wall, passing through the point ($x = 0$, $y = 0.001$ m) are extracted and shown in the third row in Fig. 3.

When the inflow consists of a pure n-heptane vapor/air mixture (Fig. 3(a)), an OSW forms as the flow interacts with the wedge. Both the temperature and pressure rise significantly upon encountering the OSW. These values continue to increase gradually as $C_7H_{16}$ fuel is consumed, followed by a sharp rise as the mass fraction of $C_7H_{16}$ approaches zero near $x = 0.05$ m. At this position, a secondary oblique detonation wave (SODW) forms near the wall. This SODW interacts with the wedge-induced OSW, facilitating the successful transition of the OSW to the ODW.

In contrast, when the inflow consists of a pure n-heptane droplet/air mixture (Fig. 3(b)), the ODW fails to form. Following the OSW, both temperature and pressure increase due to shock compression. Subsequently, the n-heptane droplets gradually break up and evaporate, as evidenced by the decrease in temperature and the rise in the n-heptane vapor mass fraction. However, no substantial increase in temperature or pressure is observed thereafter, suggesting that combustion does not occur.

To successfully form an ODW under the pure n-heptane droplet/air mixture inflow, a liquid n-heptane transverse jet with a jet-inflow momentum ratio $J = 1$ is introduced at $x = 0.05$ m, as shown in Fig. 3(c). Similar to the case in Fig. 3(b), temperature and pressure increase as the flow passes through the OSW. Subsequently, the temperature gradually decreases to around 1200 K, and the mass fraction of n-heptane vapor increases. These parameters remain stable until reaching the position of the transverse jet, which disrupts the incoming flow, leading to the formation of an OSW attached to the wedge. At the position of this jet-induced OSW, both temperature and pressure experience a sharp increase, with temperature rising to above 3000 K, accompanied by a sudden drop in the n-heptane vapor mass fraction, which reaches zero. This indicates that combustion occurs immediately after the OSW, resulting in the formation of a jet-induced ODW. This jet-induced ODW interacts with the wedge-induced OSW, directly transitioning it into an ODW.

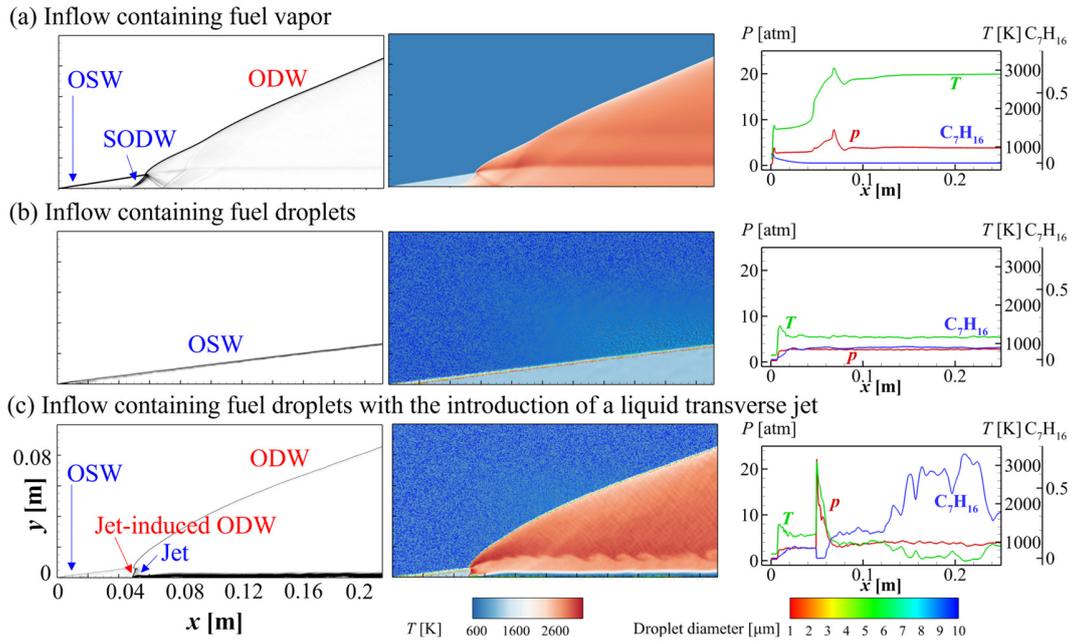

Fig. 3 Flow field for three representative cases: (a) n-heptane vapor/air inflow without a transverse jet, (b) n-heptane droplets/air inflow without a transverse jet, and (c) n-heptane droplets/air inflow with a liquid n-heptane transverse jet of $J = 1$. The first row displays numerical Schlieren images, the second row shows temperature contours and droplet distributions, and the third row illustrates flow field parameter curves along streamlines passing through the point ($x = 0$, $y = 0.001$ m) near the wall. SODW: Secondary oblique detonation wave. The droplet sizes are magnified for clarity.

### 3.2 Effects of jet-inflow momentum ratios

To investigate the influence of liquid n-heptane jet-inflow momentum ratios, three additional ratios—$J = 0.1$, 0.5, and 2.0—are considered. Figure 4 illustrates the flow fields corresponding to these jet-inflow momentum ratios. The first row presents the numerical Schlieren graphs, while the second row depicts the temperature contours and droplet distribution. Note that droplet sizes are enlarged for better visualization.

For the case of $J = 0.1$, the flame front following the OSW fluctuates around $x = 0.1$ m, and the flow field at $t = 0.1$ ms is depicted in Fig. 4(a). These figures demonstrate that the transverse jet generates a weaker OSW than the $J = 1$ case (Fig. 3(c)), with a smaller shock wave angle. This jet-induced OSW is not coupled with combustion, as no significant temperature increase or combustion after it. Subsequently, the jet-induced OSW interacts with the wedge-induced OSW, leading to a slightly stronger OSW that smoothly transitions into an ODW around $x = 0.08$ m.

To further illustrate the unsteady behavior of the ODW, Fig. 5 shows the temperature contours with droplet distribution at different times for the case with $J = 0.1$. As shown in the figure, the flow field can be divided into three regions. The first region, located below point A (the intersection of the wedge-induced OSW and jet-induced OSW), is where the unstable behavior primarily occurs. In this region, the combustion wave propagates both forward and backward over time. The second region extends from point A to point B, where the weak ODW is formed. This region also serves as the transitional zone to the main ODW. The third region, above point B, is where the main ODW is formed and remains stable.

For the case of $J = 0.5$, the wave structures are stable, as illustrated in Fig. 4(b). The jet-induced OSW near the wedge exhibits an intermediate strength between those observed for $J = 0.1$ and $J = 1$. Unlike the $J = 0.1$ case, the jet-induced OSW is followed by a high-temperature region, indicating coupling with combustion and, thus, the formation of a jet-induced ODW.

When the momentum ratio increases to $J = 2$, as shown in Fig. 4(c), the jet induces a significantly stronger shock wave compared to the $J = 1$ case. This shock wave detaches from the jet and is located approximately at $x = 0.04$ m, accompanied by a high-temperature post-shock region, classifying it as a detonation wave. The detonation wave undergoes Mach reflection at the wedge. Consequently, the total pressure loss caused by the liquid transverse jet may be substantial.

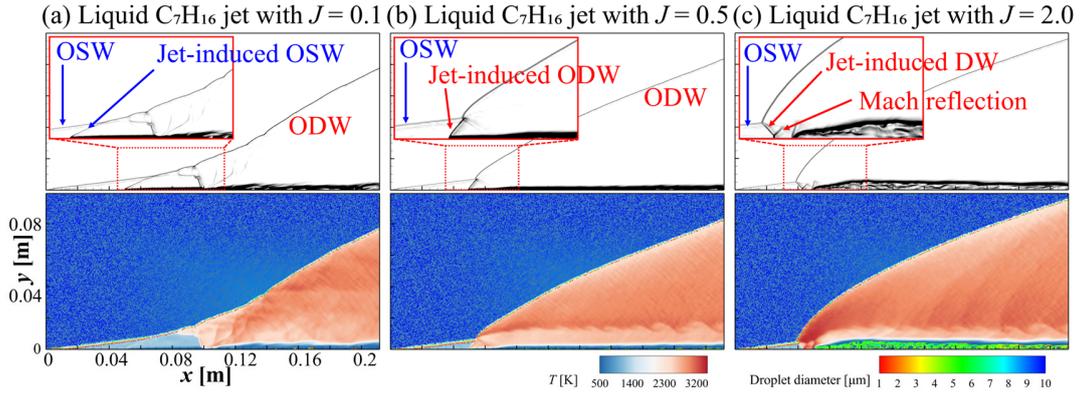

Fig. 4 Flow fields for cases with a liquid n-heptane transverse jet but different $J$ values: (a) $J = 0.1$, (b) $J = 0.5$, (c) $J = 2.0$. The first row presents the numerical Schlieren graphs, while the second row depicts the temperature contours and droplet distribution.

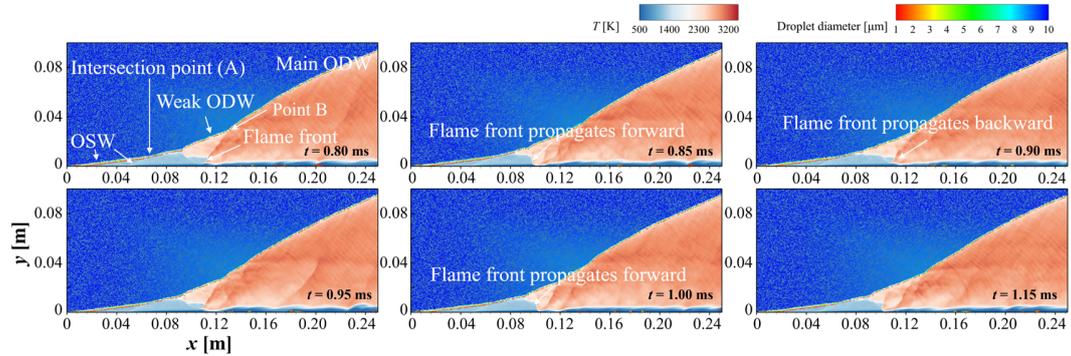

Fig. 5 Temperature contours with droplet distribution at different times for the case with the introduction of a n-heptane liquid transverse jet at $J = 0.1$.

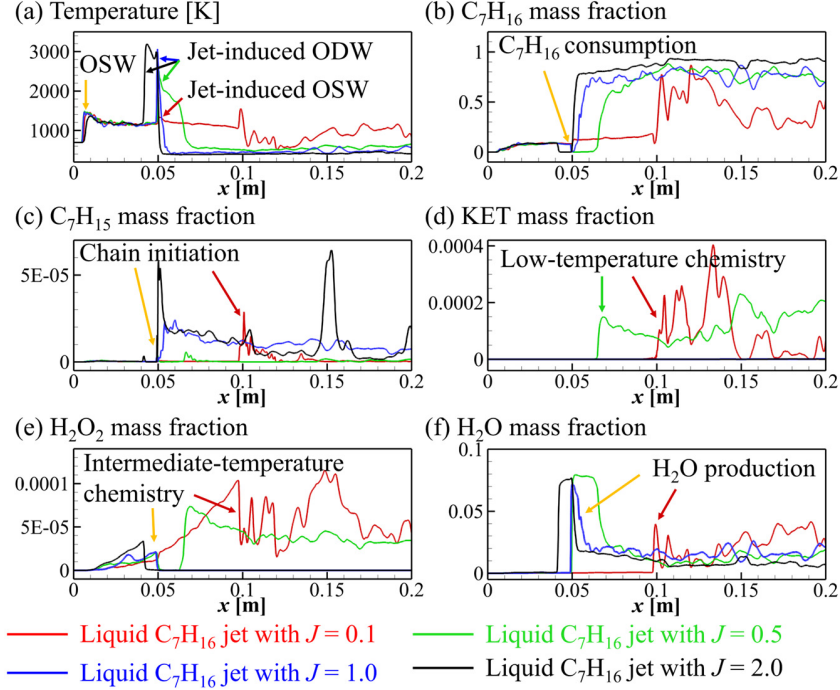

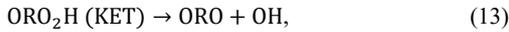
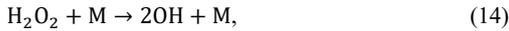

Fig. 6 Flow parameters and chemical species along streamlines passing through the same point ($x$ = 0.1 m, $y$ = 0.001 m) near the wall for cases with a liquid n-heptane transverse jet under varying jet-inflow momentum ratios.

### 3.3 Chemical kinetics analysis at different jet-inflow momentum ratios

To further investigate the flow structures and formation mechanisms under different jet-inflow momentum ratios, Figure 6 presents key flow parameters and chemical species along the streamlines passing through the same point ($x$ = 0, $y$ = 0.001 m) near the wall for four cases with varying momentum ratios. These streamlines pass through the OSW region before entering the combustion zone, providing valuable insights into the formation of combustion and the transition from OSW to ODW. Several chemical species are involved in three important reactions, including the chain initiation and two key chain-branching reactions. A chain-branching reaction is the dissociation of keto-heptyl peroxide (KET):

$$ORO_2H \text{ (KET)} \rightarrow ORO + OH, \qquad (13)$$

which has relatively high activation energy, and the buildup of KET signals low-temperature chemistry. The other chain-branching reaction is the decomposition of $H_2O_2$,

$$H_2O_2 + M \rightarrow 2OH + M, \qquad (14)$$

where the decomposition of $H_2O_2$ represents intermediate-temperature chemistry.

As shown in Fig. 6(a), these four streamlines pass through the wedge-induced OSW at approximately $x$ = 0.005 m. Downstream of the OSW, all streamlines show an increase in temperature and n-heptane vapor mass fraction, driven by the breakup and evaporation of n-heptane droplets. However, no significant reactions occur until approximately $x$ = 0.04 m, as evidenced by the mass fractions of the chain initiation product $C_7H_{15}$ and the heat release reaction product $H_2O$ remaining zero, as shown in Fig. 6(c) and (f). Beyond this point, the flow begins to diverge depending on $J$.

For the case with $J$ = 0.1, the temperature increases only slightly at $x$ = 0.05 m, as indicated by the red line, and no $H_2O$ is produced, so no detonation occurs. At approximately $x$ = 0.1 m, the n-heptane mass fraction increases due to the atomization of the n-heptane jet. The temperature then drops to around 700 K due to evaporation. At this stage, KET accumulates, $H_2O_2$ decreases, and $C_7H_{15}$ and $H_2O$ are produced, as shown in Fig. 6(c)-(f), signaling the onset of chain initiation and the coexistence of low- and intermediate-temperature chemistry.

In contrast, the n-heptane mass fraction suddenly drops to zero for the cases with $J$ = 0.5 and $J$ = 1.0 at $x$ = 0.05 m, accompanied by a peak in $H_2O$ production, signaling the onset of detonation. The ignition position for the case with $J$ = 2.0 is shifted to a shorter distance at $x$ = 0.04 m due to the strong obstacle effect of the jet; however, the trend remains consistent with the other two cases. Between $x = 0.05$ m and $x = 0.06$ m, the n-heptane mass fraction increases, while the temperature decreases to approximately 500 K due to low-temperature fuel and atomization. Consequently, $H_2O$ does not show a significant increase after $x = 0.05$ m, suggesting that substantial combustion does not occur near the wall in these three cases.

The comparison reveals that the formation mechanisms vary with different momentum ratios. For $J = 0.1$, the jet-induced shock wave is too weak to form a detonation wave directly. However, the transverse jet increases the n-heptane mass fraction and enhances low- and intermediate-temperature chemical reactions, thereby promoting the transition to detonation. This mechanism is schematized in Fig. 10 (a). In contrast, for the cases with higher momentum ratios ($J = 0.5$, 1.0, and 2.0), the jet-induced OSW is sufficiently strong to initiate detonation near the jet. However, due to stronger atomization and heat absorption, the temperature remains low near the wall, where the n-heptane mass fraction is high, and these transverse jets do not participate in combustion.

To clearly delineate these two regimes with respect to the momentum ratio, the cases with $J = 0.2$, 0.3, and 0.4 are presented and illustrated in Fig. A1 in the Appendix. It is evident that the flow fields for $J = 0.2$ and 0.3 are similar to that of $J = 0.1$, where the jet is not strong enough to directly form a detonation wave, and only a shock wave is present near the jet. In contrast, the flow field for $J = 0.4$ is similar to that of $J = 0.5$, where the jet is strong enough to directly form an ODW. These cases provide a more detailed distinction between the two ignition mechanisms. The comparison also demonstrates that $J = 0.4$ is adequate for the successful formation and stabilization of the ODW under these inflow conditions. Therefore, in different scenarios, it is crucial to identify the minimal $J$ to achieve successful formation while minimizing total pressure loss, as a large $J$ can lead to significant total pressure losses and fuel waste.

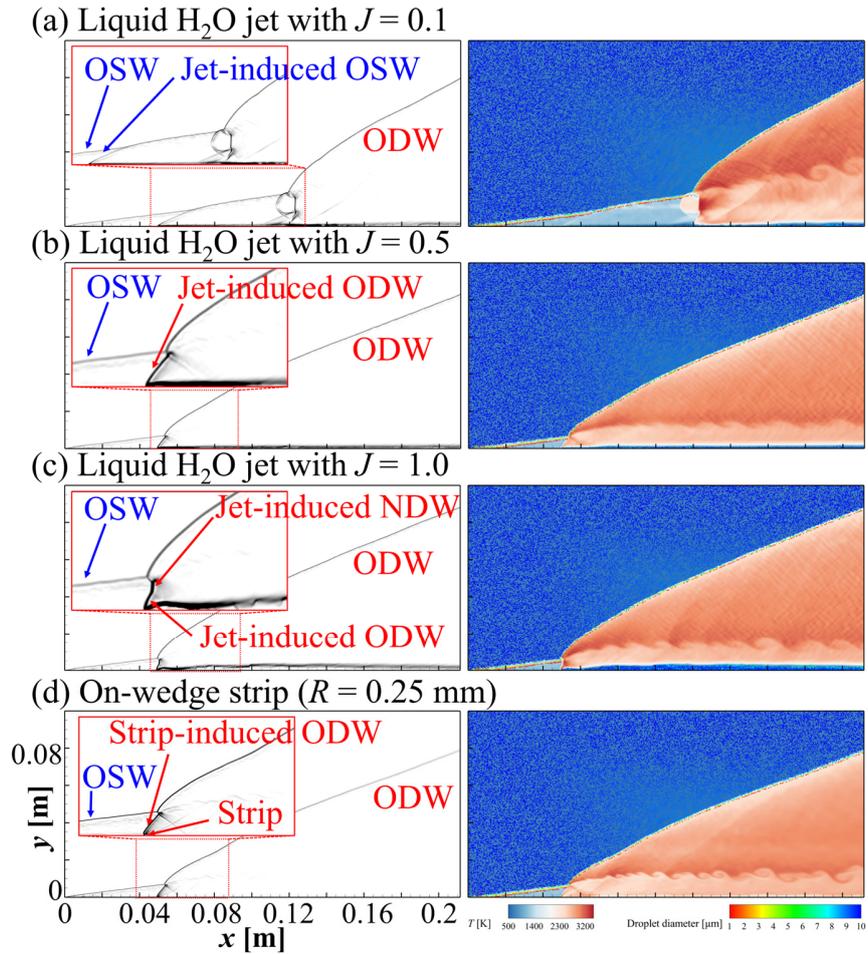

Fig. 7 Flow fields for cases with liquid $H_2O$ transverse jets at momentum ratios of (a) 0.1, (b) 0.5, (c) 1.0, and (d) a case with an on-wedge strip with a radius of 0.25 mm.

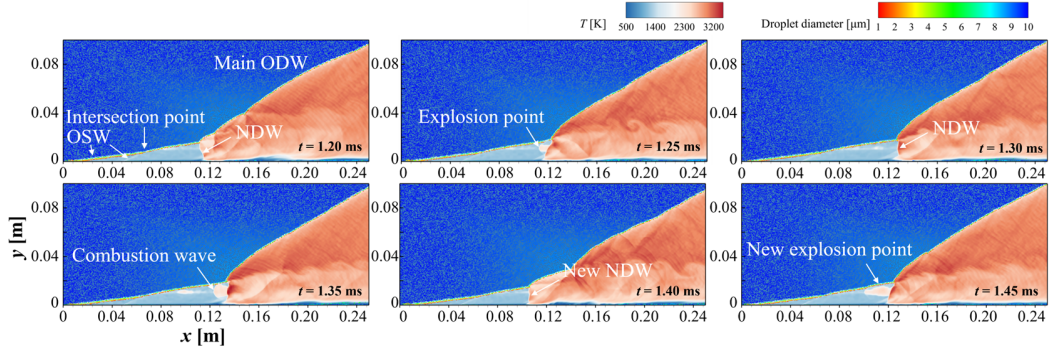

Fig. 8 Temperature contours with droplet distribution at different times for the case with the introduction of a water liquid transverse jet at J = 0.1.

### 3.4 Differences in ODW formation using non-reactive transverse liquid jets and an on-wedge strip

To investigate the formation mechanism upon introducing the liquid transverse jet, this section examines the differences and underlying mechanisms associated with the use of non-reactive transverse liquid jets and on-wedge strips in forming ODWs. These are then compared to the cases involving reactive liquid jets discussed in the previous section. The analysis includes cases with liquid $H_2O$ jets of varying momentum ratios, as well as a case involving an on-wedge strip with a diameter of 0.5 mm, which matches the diameter of the jet injector.

Figure 7(a) illustrates the flow field for the $H_2O$ transverse jet with a momentum ratio of 0.1. This flow field exhibits instability, with the flame front fluctuating near $x = 0.12$ m. Unlike the case with the $C_7H_{16}$ transverse jet at the same momentum ratio, the flow structure differs significantly. The figure reveals that at approximately $x = 0.05$ m, the $H_2O$ jet also induces an OSW. Near this jet-induced OSW, the temperature does not significantly increase, indicating the absence of combustion. This OSW interacts with the wedge-induced OSW at around $x = 0.07$ m, resulting in a slightly stronger OSW characterized by a larger shock wave angle. Subsequently, the OSW transitions to an ODW at around $x = 0.12$ m. Compared to the case with the $C_7H_{16}$ jet, the transition length is more extended.

To further explore the unsteady behavior, Figure 8 presents the temperature contours with droplet distribution at various times for the case with $J = 0.1$ and a water liquid transverse jet. In contrast to the case with an n-heptane jet, the water jet exhibits a more unstable flow field. At $t = 1.2$ ms, a normal detonation wave (NDW) forms near the wall. Subsequently, an explosion point may emerge from this NDW, only to disappear. However, at $t = 1.35$ ms, a combustion wave forms, propagating forward and generating a new NDW at $t = 1.4$ ms. This marks the end of the cycle, after which a new explosion point appears at $t = 1.45$ ms. Such unsteady behavior aligns with the findings of Teng et al [27].

For $J = 0.5$ and 1.0 (Fig. 7(b) and (c)), the $H_2O$ and $C_7H_{16}$ jets produce nearly identical flow fields. In both cases, the jet-induced OSW is strong enough to initiate detonation, forming a jet-induced ODW. This jet-induced ODW interacts with the wedge-induced OSW, causing it to transition into an ODW. Thus, the jet forms an ODW near and far from the wedge. Despite $H_2O$ being non-reactive, the flow structures mirror those of $C_7H_{16}$, highlighting that at higher momentum ratios, the jet's blocking effect dominates, generating a strong OSW and facilitating the transition. To explore this mechanism further, a case with an on-wedge strip matching the jet's position and diameter is analyzed (Fig. 7(d)). The resulting flow resembles the momentum ratio 0.5 case, where the strip induces an OSW, leading to a wedge-induced ODW.

### 3.5 Mechanisms of ODW formation with reactive and non-reactive transverse liquid jets, and an on-wedge strip at high momentum ratios

To analyze the determining formation mechanism in detail at a large momentum ratio, three cases are selected for comparison: the case with a $C_7H_{16}$ jet at $J = 0.5$, the case with an $H_2O$ jet at $J = 0.5$, and the case with an on-wedge strip. Key flow and chemical parameters along streamlines passing through the same point ($x = 0.1$ m, $y = 0.001$ m) near the wall are extracted and plotted in Fig. 9.

These three streamlines pass through the wedge-induced OSW at approximately $x = 0.005$ m, where temperature rises sharply, as illustrated in Fig. 9 (a). After this OSW, the temperature along the streamlines gradually decreases while the $C_7H_{16}$ vapor mass fraction gradually increases until around $x = 0.015$ m due to the atomization of n-heptane droplets. In the region from $x = 0.015$ m to 0.05 m, no significant combustion occurs, as indicated by the nearly zero mass fractions of $C_7H_{15}$, OH, and $H_2O$ as shown in Fig. 9(c), (f), and (g). At $x = 0.05$ m, the streamlines pass through the shock wave induced by the transverse jet or on-wedge strip. At this point, the $C_7H_{16}$ and $H_2O_2$ mass fraction for all three cases rapidly drops to zero, accompanied by a peak in OH, as well as sharp increases in temperature and $H_2O$ mass fraction, along with significant heat release, as shown in Fig. 9(h), indicating the onset of detonation.

The behavior of these streamlines diverges only after $x = 0.05$ m. For the cases involving the $H_2O$ jet and on-wedge strip, as depicted by the green and blue lines, all parameters, except for the $H_2O$ mass fraction, temperature, and heat release rate, follow similar trends beyond $x = 0.05$ m. Specifically, for the case corresponding to the green line, which represents the liquid $H_2O$ jet, since the $H_2O$ jet does not participate in the reaction, there is no significant temperature rise after $x = 0.05$ m, nor is there any accumulation of KET, consumption of $H_2O_2$, or production of OH. This indicates the absence of low and intermediate-temperature chemical reactions beyond this point. Although the $H_2O$ mass fraction increases, this is due to the evaporation of the water jet. For the case corresponding to the blue line, which represents the on-wedge strip, the temperature and heat release rate remain relatively high after $x = 0.05$ m due to the absence of a low-temperature jet. However, there is no further production of $H_2O$, indicating that no reaction occurs beyond this point. The observed similarity suggests that the $H_2O$ jet plays a role similar to the on-wedge strip in forming the ODW, primarily acting as a flow obstruction and inducing the shock wave.

The behavior of the case with the $C_7H_{16}$ jet diverges after $x = 0.05$ m, as illustrated by the red line in Fig. 9. Following the atomization of the $C_7H_{16}$ jet, the $C_7H_{16}$ mass fraction increases beyond this point, accompanied by a rise in $C_7H_{15}$ and the accumulation of KET, signaling the initiation of low-temperature chemistry. However, this occurs past the transition point and does not contribute to the formation of the ODW. Furthermore, there is no significant consumption of $H_2O_2$, nor is there any production of $H_2O$, indicating the absence of intermediate-temperature chemical reactions and heat release processes. The above analysis demonstrates that for a larger inflow-jet momentum ratio, the formation mechanism of the ODW is primarily governed by the physical effects of the jet rather than its chemical effects, as schematized in Fig. 10 (b).

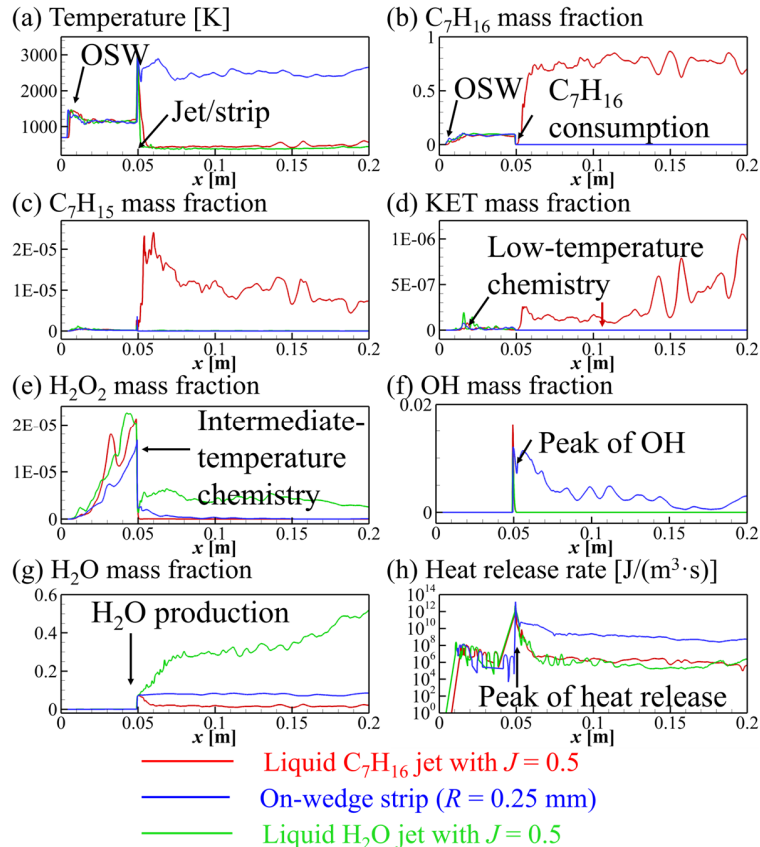

Fig. 9 Key flow and chemical parameters along streamlines passing through the same point ($x = 0.1$ m, $y = 0.001$ m) near the wall for three different ODW formation methods: red lines represent the liquid $C_7H_{16}$ jet, blue lines represent the on-wedge strip, and green lines represent the liquid $H_2O$ jet.

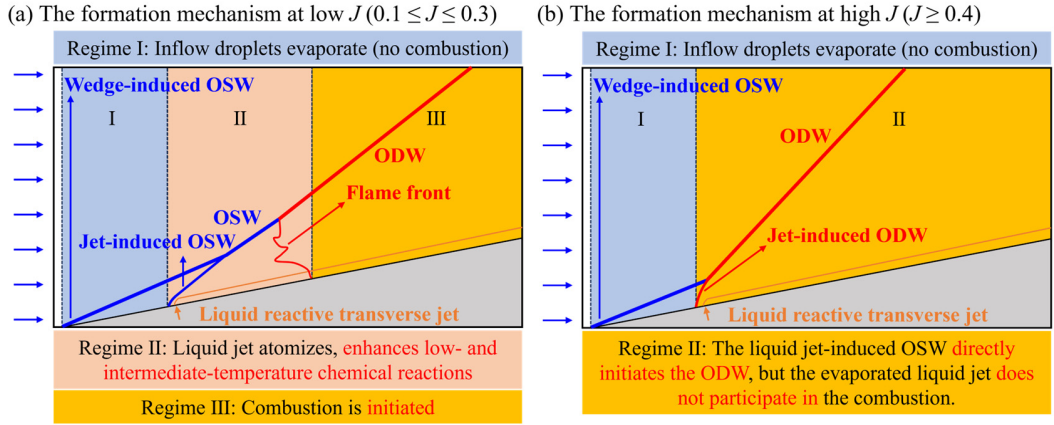

Fig. 10 Schematics of the ODW formation mechanisms induced by a liquid n-heptane transverse jet at different jet-inflow momentum ratios: (a) low momentum ratio ($0.1 \leq J \leq 0.3$), (b) high momentum ratio ($J \geq 0.4$).

## 4. Conclusions

In this computational study, using an in-house two-phase supersonic reactive flow solver based on the *rhocentralfoam* framework in OpenFOAM V7, the formation of liquid-jet-induced ODW in a liquid-fueled ODWE was realized for the first time and analyzed in detail. The results highlight the critical role of liquid transverse jets in enabling successful ODW formation in conditions where detonation would otherwise fail.

Analysis of jet-inflow momentum ratios revealed two distinct mechanisms for ODW formation. At lower momentum ratios, the jet-induced OSW is too weak to directly generate an ODW. However, the atomized n-heptane jet enhances low- and intermediate-temperature chemical reactions by increasing the fuel mass fraction, thereby promoting the transition to combustion near the wall and the conversion of wedge-induced OSW to ODW. Consequently, the configuration with the water transverse jet exhibits a longer transition length. At higher momentum ratios, the jet-induced OSW is sufficiently strong to directly initiate detonation via intermediate-temperature chemistry. In this scenario, the jet primarily serves as a physical obstacle, generating a strong jet-induced shock wave without directly participating in combustion. Comparative studies with non-reactive jets and on-wedge strip configurations further corroborate this finding, demonstrating that intermediate-temperature chemistry and heat release predominantly occur at the jet or strip locations for all ignition control settings.

Future research should investigate the impact of various jet breakup models on droplet distribution and penetration height, given their significant influence on jet behavior and subsequent detonation processes. Additionally, exploring jet-induced ODW formation in confined combustors is essential to address critical factors such as fuel loss, thrust potential, and formation control mechanisms. These considerations are vital for advancing the development of liquid-fueled ODWE technology.

## CrediT authorship contribution statement

**Wenhao Wang**: Writing – original draft, Visualization, Validation, Methodology, Investigation, Data curation. **Zongmin Hu**: Writing – review & editing, Supervision, Resources, Funding acquisition. **Peng Zhang**: Writing – review & editing, Supervision, Resources, Project administration, Funding acquisition, Conceptualization.

## Declaration of competing interest

The authors declare that they have no known competing financial interests or personal relationships that could have appeared to influence the work reported in this paper.

## Acknowledgments

This work was supported by the National Natural Science Foundation of China (Grant No. 52176134 and 12172365). The work at the City University of Hong Kong was additionally supported by grants from the Research Grants Council of the Hong Kong Special Administrative Region, China (Project No. CityU 15222421 and CityU 15218820).

## Supplementary material

The nomenclature table is provided in the Supplementary Material. Larger, higher-resolution images are also included in the Supplementary Material for improved clarity and detail, with some rearranged to ensure visibility while preserving the displayed information.

## Appendix

The appendix presents supplementary flow field structures under varying momentum ratios ($J$) with the introduction of a transverse n-heptane jet, as illustrated in Fig. A1. For $J$ = 0.2 and $J$ = 0.3, the configurations closely resemble that of $J$ = 0.1 depicted in Fig. 4(a). In these cases, the jet-induced shock wave near the wall lacks sufficient strength to directly initiate an ODW; instead, a combustion wave develops after a certain propagation distance. Furthermore, due to the intersection between the jet-induced and wedge-induced OSWs, the latter transitions smoothly into the ODW. At $J$ = 0.4, the flow field structure resembles that at $J$ = 0.5, shown in Fig. 4(b), where the jet directly initiates an ODW and facilitates the transition of the wedge-induced OSW into the ODW.

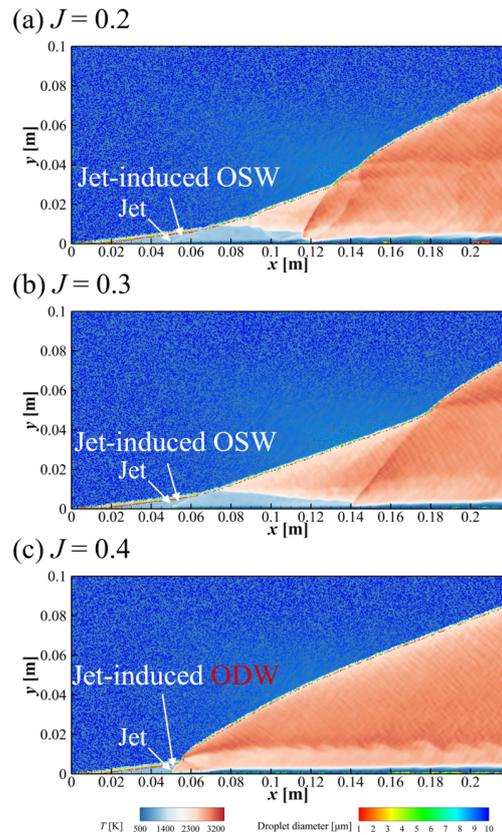

Fig. A1 Flow fields for cases with a liquid n-heptane transverse jet but different $J$ values: (a) $J$ = 0.2, (b) $J$ = 0.3, (c) $J$ = 0.4. The droplet sizes are magnified for clarity.

## References

[1] R. Dunlap, R.L. Brehm, J.A. Nicholls, A Preliminary Study of the Application of Steady-State Detonative Combustion to a Reaction Engine, J. Jet Propuls. 28 (1958) 451-456.
[2] D.T. Pratt, J.W. Humphrey, D.E. Glenn, Morphology of standing oblique detonation waves, J. Propul. Power 7 (1991) 837-845.
[3] Y. Fang, Z. Zhang, Z. Hu, X. Deng, Initiation of oblique detonation waves induced by a blunt wedge in stoichiometric hydrogen-air mixtures, Aerosp. Sci. Technol. 92 (2019) 676-684.
[4] K. Kailasanath, Review of propulsion applications of detonation waves, AIAA J. 38 (2000) 1698-1708.
[5] E.M. Braun, F.K. Lu, D.R. Wilson, J.A. Camberos, Airbreathing rotating detonation wave engine cycle analysis, Aerosp. Sci. Technol. 27 (2013) 201-208.
[6] Z. Jiang, Standing oblique detonation for hypersonic propulsion: A review, Prog. Aerosp. Sci. 143 (2023) 100955.
[7] Z. Zhang, C. Wen, W. Zhang, Y. Liu, Z. Jiang, Formation of stabilized oblique detonation waves in a combustor, Combust. Flame 223 (2021) 423-436.
[8] Z. Xiao, C. Zhang, S. Huang, S. Zhang, X. Tan, Z. Lian, J.-J. Zou, X. Zhang, G. Li, D. Wang, A comprehensive review on steam reforming of liquid hydrocarbon fuels: Research advances and Prospects, Fuel 368 (2024) 131596.
[9] X. Han, Y. Liu, Z. Zhang, W. Zhang, C. Yuan, G. Han, Z. Jiang, Experimental demonstration of forced initiation of kerosene oblique detonation by an on-wedge trip in an ODE model, Combust. Flame 258 (2023) 113102.
[10] Z. Zhang, C. Wen, C. Yuan, Y. Liu, G. Han, C. Wang, Z. Jiang, An experimental study of formation of stabilized oblique detonation waves in a combustor, Combust. Flame 237 (2022) 111868.
[11] X. Han, R. Qiu, Y. You, Flow characteristics and propulsive performance of oblique detonation waves induced by a transverse jet, Phys. Fluids 36 (2024).
[12] Y. Xin, J. Shang, G. Xiang, Q. Wang, Investigation on Accelerated Initiation of Oblique Detonation Wave Induced by Laser-Heating Hot-Spot, Aerospace, 2024.
[13] J. Sun, P. Yang, B. Tian, Z. Chen, Evolution and Control of Oblique Detonation Wave Structure in Unsteady Inflow, AIAA J. 61 (2023) 4808-4820.
[14] J. Sun, P. Yang, B. Tian, Z. Chen, Effects of wedge-angle change on the evolution of oblique detonation wave structure, Phys. Fluids 34 (2022).
[15] Q. Qin, X. Zhang, A novel method for trigger location control of the oblique detonation wave by a modified wedge, Combust. Flame 197 (2018) 65-77.
[16] A. Wang, J. Bian, H. Teng, Numerical study on initiation of oblique detonation wave by hot jet, Appl. Therm. Eng. 213 (2022) 118679.
[17] H. Li, J. Li, C. Xiong, W. Fan, L. Zhao, W. Han, Investigation of hot jet on active control of oblique detonation waves, Chin. J. Aeronaut. 33 (2020) 861-869.
[18] Q. Qin, X. Zhang, Study on the initiation characteristics of the oblique detonation wave by a co-flow hot jet, Acta Astronaut. 177 (2020) 86-95.
[19] J. Yao, Z. Lin, Numerical investigation of jet-wedge combinatorial initiation for oblique detonation wave in supersonic premixed mixture, Phys. Fluids 35 (2023).
[20] J. Fan, Y. Zhang, G. Xiang, Y. Feng, Numerical investigation of sweeping jet actuator on oblique detonation, Combust. Flame 268 (2024) 113622.
[21] Z. Ren, B. Wang, G. Xiang, L. Zheng, Effect of the multiphase composition in a premixed fuel–air stream

[21] on wedge-induced oblique detonation stabilisation, J. Fluid Mech. 846 (2018) 411-427.

[22] Z. Ren, B. Wang, G. Xiang, L. Zheng, Numerical analysis of wedge-induced oblique detonations in two-phase kerosene–air mixtures, Proc. Combust. Inst. 37 (2019) 3627-3635.

[23] Z. Ren, B. Wang, L. Zheng, Wedge-induced oblique detonation waves in supersonic kerosene-air premixing flows with oscillating pressure, Aerosp. Sci. Technol. 110 (2021) 106472.

[24] Z. Huang, M. Zhao, Y. Xu, G. Li, H. Zhang, Eulerian-Lagrangian modelling of detonative combustion in two-phase gas-droplet mixtures with OpenFOAM: Validations and verifications, Fuel 286 (2021) 119402.

[25] H. Guo, Y. Xu, S. Li, H. Zhang, On the evolutions of induction zone structure in wedge-stabilized oblique detonation with water mist flows, Combust. Flame 241 (2022) 112122.

[26] H. Guo, Y. Xu, H. Zheng, H. Zhang, Ignition limit and shock-to-detonation transition mode of n-heptane/air mixture in high-speed wedge flows, Proc. Combust. Inst. 39 (2023) 4771-4780.

[27] H. Teng, C. Tian, P. Yang, M. Zhao, Effect of droplet diameter on oblique detonations with partially pre-vaporized n–heptane sprays, Combust. Flame 258 (2023) 113062.

[28] H. Teng, C. Tian, Y. Zhang, L. Zhou, H.D. Ng, Morphology of oblique detonation waves in a stoichiometric hydrogen–air mixture, J. Fluid Mech. 913 (2021) A1.

[29] W. Wang, Z. Hu, P. Zhang, Computational investigation on the formation of liquid-fueled oblique detonation waves, Combust. Flame 271 (2025) 113839.

[30] S. Liu, J.C. Hewson, J.H. Chen, H. Pitsch, Effects of strain rate on high-pressure nonpremixed n-heptane autoignition in counterflow, Combust. Flame 137 (2004) 320-339.

[31] P. Dai, Z. Chen, X. Gan, Autoignition and detonation development induced by a hot spot in fuel-lean and CO2 diluted n-heptane/air mixtures, Combust. Flame 201 (2019) 208-214.

[32] M. Zhao, H. Zhang, Rotating detonative combustion in partially pre-vaporized dilute n-heptane sprays: Droplet size and equivalence ratio effects, Fuel 304 (2021) 121481.

[33] M. Zhao, Z. Ren, H. Zhang, Pulsating detonative combustion in n-heptane/air mixtures under off-stoichiometric conditions, Combust. Flame 226 (2021) 285-301.

[34] Q. Meng, M. Zhao, Y. Xu, L. Zhang, H. Zhang, Structure and dynamics of spray detonation in n-heptane droplet/vapor/air mixtures, Combust. Flame 249 (2023) 112603.

[35] C. Tian, H. Teng, B. Shi, P. Yang, K. Wang, M. Zhao, Propagation instabilities of the oblique detonation wave in partially prevaporized n-heptane sprays, J. Fluid Mech. 984 (2024) A16.

[36] B. Abramzon, W.A. Sirignano, Droplet vaporization model for spray combustion calculations, Int. J. Heat Mass Transfer 32 (1989) 1605-1618.

[37] R.H. Perry, D.W. Green, J.O. Maloney, Perry's Chemical Engineers' Handbook, 2007.

[38] A.B. Liu, D. Mather, R.D. Reitz, Modeling the effects of drop drag and breakup on fuel sprays, SAE Trans., (1993) 83-95.

[39] W.E. Ranz, Evaporation from drops: Part I, Chem. Eng. Prog. 48 (1952) 141.

[40] M. Pilch, C. Erdman, Use of breakup time data and velocity history data to predict the maximum size of stable fragments for acceleration-induced breakup of a liquid drop, Int. J. Multiphase Flow 13 (1987) 741-757.

[41] H. Guo, Y. Sun, R. Zhu, S. Wang, M. Zhao, B. Shi, X. Hou, Inhibition of the oblique detonation wave detachment in two-phase n-heptane/air mixtures, Combust. Flame 272 (2025) 113843.

[42] Q. Meng, Z. Wang, On the direct initiation in n-heptane mists considering droplet fragmentation, Phys. Fluids 37 (2025).

[43] C.T. Crowe, M.P. Sharma, D.E. Stock, The Particle-Source-In Cell (PSI-CELL) Model for Gas-Droplet Flows, J. Fluids Eng. 99 (1977) 325-332.

[44] Z. Zhang, Z. Jiang, Numerical Investigation of Transverse-Jet-Assisted Initiation of Oblique Detonation Waves in a Combustor, Aerospace 10 (2023) 1033.

[45] Z. Huang, Y. Xu, S. Li, Q. Meng, H. Zhang, Detailed simulations of one-dimensional detonation propagation in dilute n-heptane spray and air mixtures, Phys. Fluids 35 (2023).

[46] H. Nomura, Y. Ujiie, H.J. Rath, J.i. Sato, M. Kono, Experimental study on high-pressure droplet evaporation using microgravity conditions, Symp. (Int.) Combust. 26 (1996) 1267-1273.

[47] Y. Zhang, Y. Fang, H.D. Ng, H. Teng, Numerical investigation on the initiation of oblique detonation waves in stoichiometric acetylene–oxygen mixtures with high argon dilution, Combust. Flame, (2019).

[48] W. Wang, M. Yang, Z. Hu, P. Zhang, A dynamic droplet breakup model for Eulerian-Lagrangian simulation of liquid-fueled detonation, Aerosp. Sci. Technol. 151 (2024) 109271.

[49] C.J. Greenshields, H.G. Weller, L. Gasparini, J.M. Reese, Implementation of semi-discrete, non-staggered central schemes in a colocated, polyhedral, finite volume framework, for high-speed viscous flows, Int. J. Numer. Methods Fluids 63 (2010) 1-21.

[50] A. Kurganov, S. Noelle, G. Petrova, Semidiscrete central-upwind schemes for hyperbolic conservation laws and Hamilton--Jacobi equations, SIAM J. Sci. Comput. 23 (2001) 707-740.